\documentclass[twocolumn, prX]{revtex4-1}
\usepackage[utf8]{inputenc}
\usepackage{graphicx}

\begin{document}

\title[Lifetime enhancement for multi-photon absorption in ISBC]{Lifetime enhancement for multi-photon absorption in intermediate band solar cells}

\author{A T Bezerra$^{1,3}$, N Studart$^{2,3}$}
\address{$^1$ Departamento de Física, Instituto de Ciências Exatas, UNIFAL, 37130-000, Alfenas, MG, Brazil}
\address{$^2$ Centro de Ci\^encias Naturais e Humanas, UFABC, 09210-580, Santo Andr\'e, SP, Brazil}
\address{$^3$ DISSE, Instituto Nacional de Ciência e Tecnologia de Nanodispositivos Semicondutores, Brazil}

%\ead{anibal.bezerra@unifal-mg.edu.br}

\begin{abstract}
A semiconductor structure consisting of two coupled quantum wells embedded into the intrinsic region of a {\it p-i-n} junction is proposed to be implemented 
as an intermediate band solar cell with ratchet state. The localized conduction subband of the right-hand side quantum well is thought as 
the intermediated band, while the excited conduction subband of the right-hand side quantum well, coupled to right-rand side one, is thought to acts
as the ratchet state. The photo-excited electron in the intermediate band can tunnel out the thin barrier separating the wells and accumulate into ratchet subband. 
This might raise the electron probability of being hit by a second photon and exiting out to the continuum, increasing solar cell current. Is presented a temporal 
rate model for describing the charge transport properties of the cell. Calculations are carried out by solving the time-dependent Schrödinger 
equation applying the time evolution operator within a pertinent choice of the non-commuting kinetic and potential operators. The efficiency in the generation of 
current is analyzed directly by studying the occupation of the subbands wells in the p-i-n junction, taking into account the injection and draining dynamic provided by
the electrical contacts connected to the cell. As a result, the efficiency in the generation of current was found to be directly correlated to the relationship between
optical generation and recombination rates regarding to the scattering to the ratchet state rate. This suggests that a good coupling between the intermediate band and 
the additional band is a key point to be analyzed when developing an efficient solar cell. 
\end{abstract}

\keywords{Intermediate Band Solar Cells; Multi-Photon Process; Quantum Well}
\pacs{1315, 9440T}
%\submitto{\JPD}
\maketitle
%\ioptwocol

\section{Introduction}

The demand for renewable energy sources has been promoting the research on semiconductor solar cells [1-12]. %~\cite{SCHO61,LUQU97,YOSH12,NODA16,BARN90,CABR13,LUQE12,ANDE95,MART06,ELBO15,PUSC16,DATA15}.
Basically, a solar cell is formed by a {\it p-i-n} junction shed with light 
in order to excite charge carriers between valence and conduction bands. The carriers are then drained by the contacts by a drift electric field, giving rise to a net electric 
current~\cite{SUPR12}. However, to a single band gap cell, the detailed balance determines the current generation to be fundamentally limited by the absorption of 
photons with energies greater than the cell's band gap~\cite{SCHO61}. To overcome such limit, 
intermediate bands solar cells has been proposed~\cite{LUQU97}. A set of electronic states, called intermediate band, is introduced within the semiconductor band gap. This
affords new pathways to the carriers~\cite{LUQU97,YOSH12,NODA16}, allowing additional carrier generation due to the extension
of the spectral response in the energy range bellow the host-semiconductor band gap~\cite{BARN90,CABR13}. Consequently, the addition of the intermediate band increases
the light-to-current conversion efficiency of the solar cell. An additional condition to ensure such an efficiency enhancement, is that intermediate band needs to be 
radiatively connected to the valence band but electrically isolated from the other bands~\cite{LUQE12}.

The use of the localized subbands of quantum wells, embedded into the intrinsic region of {\it p-i-n} solar cells, are promising candidates to be used as the intermediate 
bands~\cite{YOSH12}. They has been suggested to effectively enhance solar cell efficiencies~\cite{BARN90,ANDE95}. However, extracting carriers from 
localized states often requires inelastic processes, as multiphoton excitation~\cite{MART06,ELBO15}. That becomes a new problem due to the low carrier lifetime into the intermediate subbands 
regarding to the interband recombination process~\cite{YOSH12}. 

Hence, structures with ratchet states has been proposed~\cite{YOSH12,NODA16,CURT16}. The ratchet are states optically- and electrically-coupled to conduction band 
and decoupled from the valence band, acting as scattering channels for electrons excited into the intermediate band.  The coupling between intermediate band and ratchet 
states, allows for excited electrons to increase their lifetimes within the intermediate band. This might increases the multiphoton transition probability 
and, consequently, the solar cell efficiency~\cite{PUSC16}. 

The use of quantum cascade-like scattering to work as the ratchet transport dynamics has been 
theoretically~\cite{CURT16} and experimentally~\cite{SUGI12} proposed. It showed to be feasible and promising for increasing the cell's efficiency. In such approach,
the energy difference between the intermediate and hatchet states are set to be in resonance with the host-semiconductor LO-phonon energy, allowing for a phonon-assisted 
scattering. This spatially shift the optically generated electron-hole pair, decreasing the recombination probability, and also increases the electron's lifetime within the 
intermediate band. However there still has a lack of information about the transport dynamics within the excitation/scattering/excitation process.

Obviously, studies of this kind of structures must rely on their quantum properties, once we are dealing with quantum devices. Therefore the analyses of such a systems under
the quantum mechanics perspective is fundamental, once we consider the quantum well subbands as the basis of the cell's operation~\cite{CABR13, DATA15,SUPR12}.

In what follows, we propose an {\it p-i-n} layered structure with two coupled quantum wells embedded into the intrinsic semiconductor region, to be implemented as
an intermediate band solar cell with ratchet state. The transport dynamics is based on the quantum cascade-like scattering at the intermediate band.
We analyze the current response of the cell by means of a rates model using the recombination and generation rates, directly
obtained through the system eigenfunctions. We find a direct correlation between the scattering rate to the ratchet state and the recombination rates, rising the solar 
cell current. However, we observe conditions that even having the scattering to the ratchet states, the resulting current could result in a decrease of the net current.

\section{Methods}

\subsection{Structure}

As discussed earlier, the major problem of using quantum well subbands as intermediate bands is the lowest excitation time for the electron at the intermediate band 
regarding to its recombination (back to valence band) time. Allowing such an electron to relax into a ratchet state might increase its lifetime and, consequently, 
the two-photon process efficiency~\cite{CURT16, PUSC16}. 

\begin{figure}
 \centering
 \includegraphics[scale=0.38]{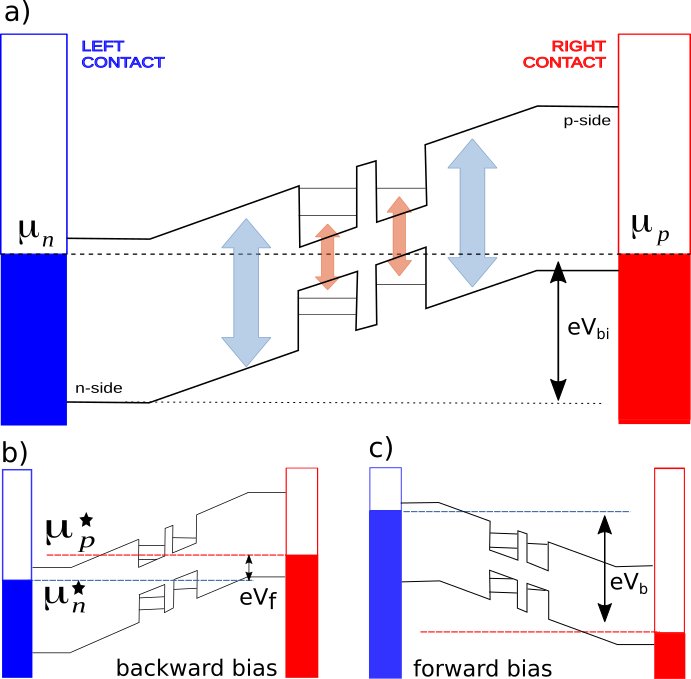}
 % IBSCPotentialProfile.jpg: 1058x794 pixel, 72dpi, 37.32x28.01 cm, bb=0 0 1058 794
 \caption{(color online) Schematics of the potential profile of the intermediate band solar cell with ratchet states. (a) Structure
 without aplying external potential, the equilibrium between doped contacts gives rise to a built-in potential $eV_{bi}$. The arrows represent the optical transitions responsible to 
 generation of photocurrent. (b) Structure under backward external bias $eV_f$. $\mu_{n(p)}^{\star}$ represent the quasi-fermi levels for conduction (valence) band. (c)
 Structure under forward external bias $eV_b$.}
 \label{fig:potentialProfile}
\end{figure}

In order to enhance the generation of current, the chosen potential profile was thought to have two quantum wells, and the work is based on dealing with 
their absorption and emission dynamics. 
As we can observe in the~\ref{fig:potentialProfile}, the left-hand side quantum well (lQW) is wider, having two confined subbands in the conduction band, while the right-hand 
side one (rQW) is thinner having only one. The intention was to use the rQW subband as the intermediate band and the lQW's excited subband as the ratchet state. 
Being the latter an excited state, the selection rules for interband transition uncouple it for the valence band ground state~\cite{BAST92}, as required~\cite{YOSH12}. 
This also enables new pathways to the intersubband absorption process hopefully increasing even more the current generation.

In details, the resulting intermediate band solar cell under analysis, was based on a layered structure within material and parameters found in the literature~\cite{NODA16}. 
The $1~\mu m$ wider intrinsic layer and the doped contacts has been
chosen to be Ga$_x$Al$_{1-x}$As, with $x=0.3$. The left- and the right-hand side quantum wells are formed by GaAs layers with widths 110~\AA~ and 40~\AA, respectively, coupled each 
other by a 80~\AA~ Ga$_x$Al$_{1-x}$As barrier, as shown in the~\ref{fig:potentialProfile}. The energy shift between the rQW ground state and the  lQW excited state was
set to be around $32$ meV, the GaAs LO-phonon energy~\cite{SING03}.

It has been chosen the drain and injection contacts, $n$ and $p$ layers, respectively, to be doped with concentrations of $N = 10^{17}~cm^{-3}$. 
Without any external electric potential applied to the device, 
at thermal equilibrium, the Fermi level is the same throughout the structure (see the bold black-dashed line in the~\ref{fig:potentialProfile}(a)).
This gives rise to a built-in potential, $V_{bi} = 4.23$~eV, by the depletion of the potential profile at intrinsic region. Then a drift field is established, 
responsible for extracting photogenerated carriers towards the drain ($n$) contact~\cite{SUPR12}. In order to restore the thermal equilibrium, the same amount of 
drained charge is 
re-injected by the injection ($p$) contact into the cell, having as a net effect the generation of a short circuit current on the external circuit.

Under external bias, the potential profile changes due to the equilibrium within the contacts' electrochemical potentials, as shown in the~\ref{fig:potentialProfile}(b) at 
backward bias condition, and in the~\ref{fig:potentialProfile}(c) at the forward bias condition. It was considered for the levels occupation the equilibrium with the 
quasi-Fermi levels established by the contacts. At forward bias condition, we do not we expect optical generation of current, since the conduction band states, 
at thermal equilibrium, are kept fulled. Under such a bias condition, current is expected only in the presence of inelastic scattering processes 
(common to a $p-i-n$ diode)~\cite{SUPR12}, not considered in the present model.

The system was modeled using uncoupled parabolic valence and conduction bands within the effective mass approximation~\cite{BAST92}. The eigenfunctions
were numerically obtained by solving the time-dependent Schrödinger equation through the Split 
Operator method, within the imaginary time evolution~\cite{DEGA10}. In the method, an initially guessed wave function is evolved in time by successive applications of 
the time-evolution operator, $\exp(-iHd\tau/\hbar)$, choosing  $d\tau = -idt$. However, the system's Hamiltonian, $H = K + U$, formed by the non-commuting kinetic ($K$) and potential 
operators ($U$), imposes an intrinsic error when applied to the time-evolution operator. The error can be easily handled by properly splitting the exponential argument. 

Choosing

\begin{equation}\label{eqSplitOperator}
 e^{-i\frac{Hdt}{\hbar}} = e^{-i\frac{Udt}{2\hbar}} e^{-i\frac{Kdt}{\hbar}} e^{-i\frac{Udt}{2\hbar}}+ O(dt^3),
\end{equation}
the error in time evolution can be arbitrarily controlled by appropriate choice of the time increment $dt$, once it is proportional to the third order of $dt$. The excited subbands
are obtained carrying out the time evolution together with the Gram-Schmidt orthonomalization scheme. Within the eigenfunctions, the subbands' energies are obtained by the mean
of the system's Hamiltonian. 

\subsection{Rates Model}

\begin{figure}
 \centering
 \includegraphics[scale=0.2]{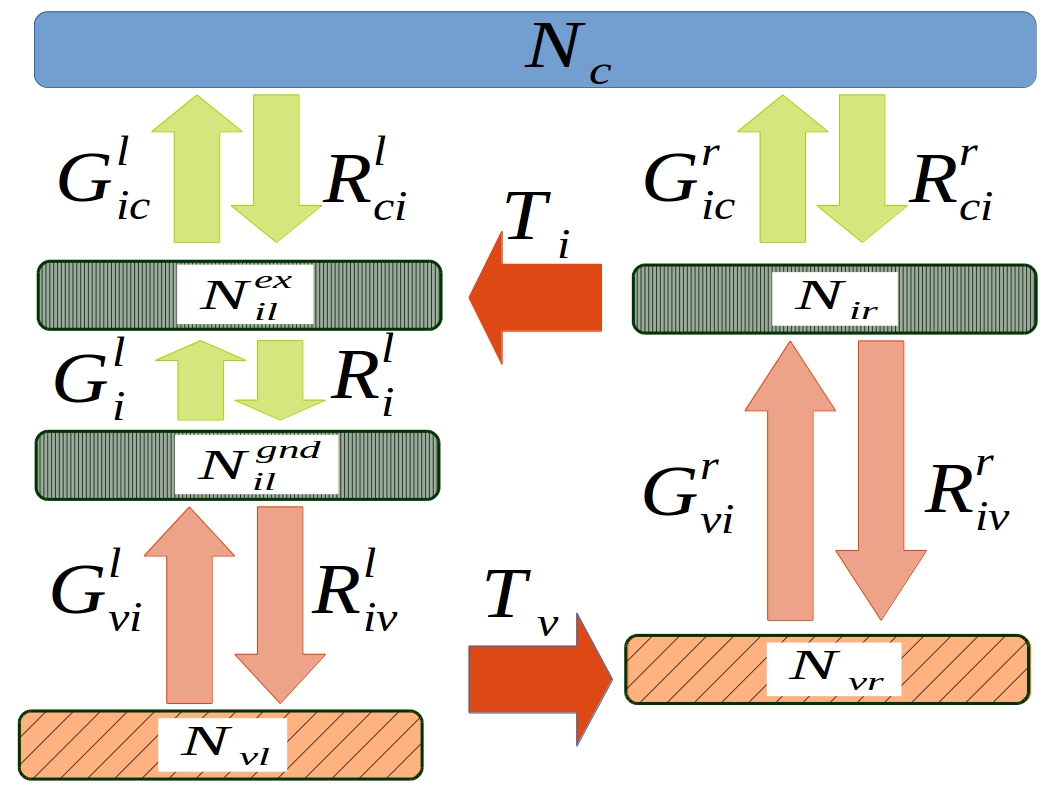}
 % levelsDynamics.jpg: 1058x794 pixel, 72dpi, 37.32x28.01 cm, bb=0 0 1058 794
 \caption{(color online) Levels schematics showing the possible transitions and their respective rates. The spaced-hatch orange boxes represent the valence levels at the 
 lQW and rQW, within carrier concentrations of $N_{vl}$ and $N_{vr}$, respectively. The closed-hatch green boxes represents the intermediate levels, within carriers 
 concentrations of $N_{ir}$, $N_{il}^{gnd}$, and $N_{il}^{ex}$, considering the left well having a ground and an excited level. $N_c$ is the conduction level 
 carrier concentration. $G(R)_{ij}^s$ is the generation (recombination) rate for the absorption between bands $i$ to $j$ at the $s$-hand side quantum well. 
 $T_{v(i)}$ is the transition rate between lQW and rQW at valence (intermediate) band.}
 \label{fig:levels}
\end{figure}

In order to model the electronic excitation and recombination dynamics, it was developed a semiclassical rates model taking into account the electronic 
subbands as  simple levels, as shown in the \ref{fig:levels}. 

The optical processes lead to changes in the carrier concentration of the levels within time. 
The spaced-hatch orange boxes in the \ref{fig:levels}, represent the valence levels at the 
lQW and rQW, within carrier concentrations of $N_{vl}$ and $N_{vr}$, respectively. The closed-hatch green boxes represent the intermediate band levels, within carriers 
concentrations of $N_{ir}$, $N_{il}^{gnd}$, and $N_{il}^{ex}$ (considering the left well having a ground and an excited level). $N_c$ is the carrier concentration of the
conduction band level. $G(R)_{ij}^s$ is the generation (recombination) rate for the absorption between bands $i$ to $j$ at the $s$-hand side quantum well. 
$T_{i(v)}$ is the transition rate between lQW and rQW at intermediate (valence) band. This rate is related to the phonon-assisted scattering process between the intermediate band and the
ratchet state. It was set as a control parameter for the simulation.

The changes of the carriers concentrations within time are

\begin{eqnarray} \label{eqRate2}
        %\dot{N}_{vl} = G_0 + R_{iv}^l N_{il}^{gnd} - (G_{vi}^l + T_v) N_{vl} \\ \label{eqRate1}
        %\dot{N}_{vr} = T_v N_{vl} + R_{iv}^r N_{ir} - G_{vi}^r N_{vr} \\ \label{eqRate2}
        \dot{N}_{ir} = G_{vi}^r N_{vr} + R_{ci}^r N_c - (G_{ic}^r + R_{iv}^r) N_{ir} - T_i N_{ir}\\ \label{eqRate3}
        \dot{N}_{il}^{ex} =  T_i N_{ir} + R_{ci}^l N_c + G_{i}^l N_{il}^{gnd} - (G_{ic}^l + R_{i}^l) N_{il}^{ex} \\ \label{eqRate4}
	\dot{N}_{il}^{gnd} = G_{vi}^l N_{vl} + R_i^l N_{il}^{ex} - (G_i^l + R_{iv}^l) N_{il}^{gnd} \\ \label{eqRate5}
	\dot{N}_c =  G_{ic}^r N_{ir} + G_{ic}^l N_{il}^{ex} - (R_{ci}^l + R_{ci}^r) N_c, \label{eqRate6}
\end{eqnarray}
where $\dot{N}_{band}$ is a short representation to $dN_{band}/{dt}$. The change with time in the electrons' concentration within a specific band.

Let's begin with the \ref{eqRate2}. At steady state condition, $\dot{N}_{ir} = 0$, so

\begin{equation}\label{eqNir}
  N_{ir} =\frac{G_{vi}^r N_{vr} + R_{ci}^r N_c}{G_{ic}^r + R_{iv}^r + T_i}. 
\end{equation}

\ref{eqRate3}, also at steady state condition ($\dot{N}_{il}^{ex} = 0$), yields

\begin{equation}\label{eqNilEx}
  N_{il}^{ex} = \frac{G_{i}^l N_{il}^{gnd} + R_{ci}^l N_c + T_i N_{ir} }{G_{ic}^l + R_{i}^l}.  
\end{equation}

We can use \ref{eqRate4} to determine the relationship between excited subband in the lQW, considering once more the steady state condition, 
$\dot{N}_{il}^{gnd} = 0$. This gives

\begin{equation}
  N_{il}^{gnd} =  \frac{G_{vi}^l N_{vl} + R_i^l N_{il}^{ex}}{G_i^l + R_{iv}^l}.
\end{equation}

Backing to the \ref{eqNilEx}, and after some simple algebra

\begin{equation}\label{eqNilexfinal}
  N_{il}^{ex} = \frac{G_{i}^lG_{vi}^lN_{vl} + (G_i^l + R_{iv}^l)(R_{ci}^l N_c + T_i N_{ir})}{\gamma},
\end{equation}
where $\gamma = (G_i^l + R_{iv}^l)( G_{ic}^l + R_{i}^l) + G_{i}^l  R_i^l$.

The current extracted from well's region was considered to be proportional to the change of the carriers concentration at conduction band, $\dot{N}_c$. 
Substituting~\ref{eqNir} and~\ref{eqNilexfinal} into~\ref{eqRate6}, yields

\begin{eqnarray}\label{eqdNcdt}
 \dot{N}_c &= \kappa N_c + \frac{ G_{ic}^l G_{i}^lG_{vi}^l}{\gamma}N_{vl} + \\  \nonumber		       
		   &\qquad {} \left[G_{ic}^r+\frac{ (G_i^l + R_{iv}^l)  G_{ic}^l T_i}{\gamma}\right]\frac{ G_{vi}^r }{G_{ic}^r + R_{iv}^r + T_i}N_{vr}  ,
\end{eqnarray}
where 

\begin{eqnarray}\label{eqKappa}
 \kappa &= R_{ci}^l\left[ \frac{(G_i^l + R_{iv}^l)G_{ic}^l }{\gamma} - 1 \right] + \\ \nonumber
        &\qquad {} R_{ci}^r\left[ \frac{(G_i^l + R_{iv}^l)G_{ic}^l T_i}{(G_{ic}^r + R_{iv}^r + T_i)\gamma} + \frac{1}{G_{ic}^r + R_{iv}^r + T_i}- 1 \right],
\end{eqnarray}
depends on the recharge of the wells via conduction band states, with rates $R_{ci}^{l(r)}$.

\subsection{Transition rates}

The photo-generation process is, mainly, a result of two interband transitions, one at the bulk region and other at the quantum wells region. However, in the latter one, 
to extract carriers from the confined subbands a second transition needs to take place ~\cite{YOSH12,NODA16}. Both the tunneling and the intersubband transition 
to continuum states, extended throughout the contacts region, can be used as such a secondary process~\cite{SUGI12,CURT16}. 
Therefore, the knowledge of the transitions rates interplay is fundamental.

We have determined the rates by means of the Fermi's golden Rule~\cite{BAST92,SING03}, using the eigenstates resulting from the Split-Operator method.
The generation rate for electrons excited from valence band to conduction band at bulk region, by photons with energy $\hbar\omega$, 
is given by~\cite{SING03}

\begin{equation}\label{eqBulkAbsorptionRate}
 G_{vc}^{bulk}(\hbar\omega) = \omega_0\frac{n_{ph}}{\hbar\omega}\left( \frac{2 p_{vc}^2}{m_0}\right)\frac{2}{3}D_{vc}(\hbar\omega),
\end{equation}
where $\omega_0 = \pi e^2 \hbar/m_0 \varepsilon$, $e$ and $m_0$ are the electron charge and mass, respectively, $\varepsilon$ is the GaAs dielectric constant, 
$n_{ph}$ is the photons' density set to unity, and $p_{vc}$ is the momentum matrix element, which can determined using the Kane approximation~\cite{BAST92}. $D_{vc}$ is the 
three-dimensional electronic density of states given by

\begin{equation}
 D_{vc}(\hbar\omega) = \sqrt{2}\frac{(m^*)^{3/2}\sqrt{\hbar\omega - E_g}}{\pi^2\hbar^3},
\end{equation}
being $m^*$ the electron's effective mass, and $E_g$ the semiconductor band gap.

At the quantum wells region, the generation rates, $G_{vi}^{l(r)}$, are obtained by changing the three-dimensional density of states to a two-dimensional one, 
taking into account the overlap of the envelope functions. So

\begin{equation}\label{eqDvcQW}
 D_{vc}^{qw}(\hbar\omega) = \frac{m^*}{\pi\hbar^2 L_{qw}}\sum_{m,n}|\left\langle \phi_v^m | \phi_c^n \right\rangle|^2 \Theta(E_{mn} - \hbar\omega),
\end{equation}
where the sum is done over the valence $\phi_v^m$ and conduction band $\phi_c^n$ eigenstates, separated in energy by $E_{mn} =E_g - E_n^c - E_m^v$. $L_{qw}$ is the well's width,
and $\Theta$ is the Heaviside step function, related to the two-dimensional density of states.

The interband generation rates, $G_{vi}^{l(r)}$ are obtained by

\begin{equation}\label{eqGvilr}
 G_{vi}^{l(r)} = \omega_0\frac{n_{ph}E_p}{\hbar\omega}\sum_{m,n} D_{vc}^{qw} f(E_m)[1-f(E_n)],
\end{equation}
where $E_p = 23$~eV is the Kane energy for GaAs~\cite{SING03}. The sum is done over the valence ({\it m}) and conduction band states ({\it n}), weighted by their occupations
given by the Fermi distributions $f\left(E_{m(n)}\right)$.  The quasi-Fermi level were determined by the contacts electrochemical potentials~\cite{SUPR12}.

The interband recombination rates  ($R_{vc}^{bulk}$, $R_{iv}^{l(r)}$) are similar to the interband generation ones. However, we might to change the density of photons to $n_{ph}+1$, 
and to replace the electronic density of states by the photons density of states $\rho(\hbar\omega) = (\hbar\omega)^2 / \pi^2 (\hbar v)^3$, where $v$ is the light velocity within the
semiconductor~\cite{SING03}.

The intersubband generation rates at the quantum wells region, $G_{ic}^{l(r)}$, are given by

\begin{equation}\label{eqQWAbsorptionRate}
 G_{ic}^{l(r)}(\hbar\omega) = \omega_0\frac{n_{ph}}{\hbar\omega L_{qw}}\sum_{i,f} |p_{if}|^2 f(E_i)[1-f(E_f)],
\end{equation}
where the sum is done over the quantum well localized subbands ({\it i,j}), again taking into account their occupations by using the Fermi distributions $f(E)$. The intersubband momentum matrix element is
$ p_{if} = -i\hbar\langle \psi_i^{l(r)}| z |\psi_f^{l(r)}\rangle / L_{qw}$, where $\psi_i^{l(r)}$ is the left (right) quantum well \textit{i-th} subband eigenfunction, 
with energy $E_i$. The intersubband recombination rate at the lQW, $R_i^l$ was considered to be constant, within a value of 10~GHz.

With the evaluated generation and recombination rates we are able, to feed the rates model and determine the level dynamics. The results are presented in the next section.

\section{Results}

In the solar cell, the electrons excited both from valence and intermediate band, towards conduction band, should be extracted from quantum wells regions by the drain contact, restoring 
the thermal equilibrium. Hence, in the model, we considered the infinity mobility regime, allowing effectively collection of charge by the drain contact ~\cite{LUQU97}.
The regime is achieved preventing electrons to relax back from continuum to the wells' subbands, choosing $R_{ci}^l = R_{ci}^r = 0$ in the~\ref{eqKappa}, so $\kappa = 0$.

To analyze the contribution of adding the left-hand side quantum well to the structure, the quantum wells were first assumed to be uncoupled from each other, 
making $T_i = 0$ in the~\ref{eqdNcdt}, yielding

\begin{equation}\label{eqdNcdt_uncoupled}
 \dot{N}_c|_{(T_i=0)} = \frac{ G_{ic}^l G_{i}^lG_{vi}^l}{\gamma}N_{vl} + \frac{G_{ic}^r G_{vi}^r }{G_{ic}^r + R_{iv}^r}N_{vr}.
\end{equation}

Clearly, when uncoupled, both the lQW and the rQW might independently contribute for changing the concentration of carriers in the conduction band. The contribution is
proportionally to their valence band concentrations, $N_{vl}$ and $N_{vr}$ weighted by their specific generation $G_{ic}^{l(r)}$, regarding to the recombination rates
$R_{iv}^{l(r)}$. The greater the recombination rate regarding to the generation one, the lesser effective is the current generation process.

We now may think in terms of a liquid current, $I_l$,
proportional to the change in the concentration of conduction carriers (times the electronic charge), when comparing the system within coupled ($T_i \neq 0$) and uncoupled ($T_i = 0$) quantum wells,

\begin{equation}
 I_{l} = e\left[\dot{N}_c |_{T_i\neq0} - \dot{N}_c |_{T_i = 0}\right] .
\end{equation}

Subtracting~\ref{eqdNcdt_uncoupled} from~\ref{eqdNcdt}, $I_l$ reads

\begin{eqnarray}\label{eqLiqdCur}
 I_{l} & = \frac{eG_{vi}^r T_i}{G_{ic}^r + R_{iv}^r + T_i} \times \\ \nonumber
                                & \qquad {} \left[ \frac{(G_i^l + R_{iv}^l)  G_{ic}^l}{\gamma} - \frac{G_{ic}^r}{G_{ic}^r + R_{iv}^r}\right].       
\end{eqnarray}

Such a liquid current was interpreted as the contribution (to the current) of adding the additional scattering channel, charging the ratchet state. It would enhances the carriers 
lifetime at the intermediate band, increasing the cell's efficiency.

\begin{figure}
 \centering
 \includegraphics[scale=0.38]{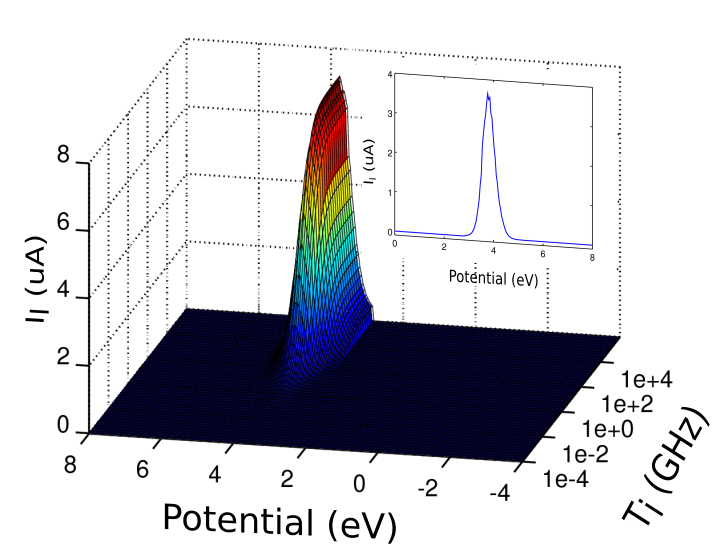}
 % TilR1d10_3D.pdf: 0x0 pixel, 300dpi, 0.00x0.00 cm, bb=
 \caption{(color online) Liquid current as a function of the potential applied to the cell and the transition rate $T_i$ between the lQW e rQW. The inset shows the profile of 
 liquid current for $T_i = 132,5$~GHz.}
 \label{fig:Til3d}
\end{figure}

In the~\ref{fig:Til3d}, we presented the liquid current, given by the~\ref{eqLiqdCur}, as a function of the transition rate $T_i$ and the 
cell's potential. The potential was varied from backward to forward biases relative to the built-in potential, uniformly spread through 
the cell's intrinsic region (see~\ref{fig:potentialProfile} (b) and (c)).  For $T_i$ greater than 1 GHz, we observed an effective enhancement of the liquid current, with a 
pronounced peak close to the built-in potential value, for potentials between 3 to 4 eV. Peak intensity increases with increasing wells' coupling rate until saturates 
for $T_i$ greater than 500 GHz. Such a behavior was understood as follows.

As given by~\ref{eqGvilr},
the interband generation rate ($G_{vi}^r$), and the interband recombination rate ($R_{iv}^r$) are determined by a conjunction of three factors - the $\Theta$ step function, 
the Fermi distributions, and the overlap between valence and intermediate subbands eigenfunctions ~\cite{SING03}.
Therefore, due to the step function, low energetic photons (regarding to the bands separation) are unable to excite electrons from valence to intermediate subband. 
The excitation of electrons by higher energetic photons is conditioned by the external bias applied to the cell (which controls the quasi-Fermi levels),
and by the overlap between the valence and intermediate band eigenstates (which decreases with increasing the energy separation from each other).

For backward bias condition, encompassing the built-in potential condition, the hole subbands are considered to be kept filled through the equilibrium 
within the injection contact, while the electron subbands are considered to be kept empty by the drain (see the quasi-Fermi levels in the~\ref{fig:potentialProfile}(b)). 
Instead, for the forward bias condition, the electron subbands are kept filled while the hole subbands are kept empty (see the quasi-Fermi levels in the~\ref{fig:potentialProfile}(c)).
In the latter condition, the current through the cell is 
non-null only in the presence of inelastic scattering processes~\cite{SUPR12}. Therefore, the quasi-Fermi levels, determined by the
potential applied to the cell, is a fundamental factor for determining both the generation and the recombination rates. 
That is directly reflected in the behavior of the cell's current, as discussed next.

As noted, the liquid current is non-null only for backward bias, as expected by the levels occupations. Analyzing~\ref{eqLiqdCur} in 
details, the first term inside the square-brackets, related to the transitions into the lQW, presents a step-like shape being close to unit for potentials 
lesser than $\sim 4$~eV, and null elsewhere. The second term, related to the transitions into the rQW, presents the same step-like fashion
whose intensity is close to unit for potentials lesser than $\sim 3$~eV, and null elsewhere. Therefore, the subtraction between such a terms determines the peak width. 
The peak edges are given by the intensities relationship between the generation and recombination processes at the lQW and rQW regions. 

The step-like profile,
discussed above, is mainly due to the step-like shape of the recombination rates  $R_{iv}^l$ and  $R_{iv}^r$. Even though the intensities of the recombination rates are
three to four orders of magnitude lower then the generation ones, the last are peaked functions with maxima for specific biases, null elsewhere. For such a biases, both
the first and the second terms inside the square-brackets of the~\ref{eqLiqdCur} are close to the unit, since the dominant terms are the generations rates. 
However, for biases greater than 3~eV (4~eV), $R_{iv}^r$ ($R_{iv}^l$) is the dominant rate, becoming the square-brackets term null. Consequently, the liquid current is 
non-null only in that specific potential range.

Then the inclusion of the lQW and its relationship with the rQW determines the presence of an enhancement in the cell's current. The peak width in the liquid current is,
therefore, controlled by subbands occupation of both wells, which determines the recombination rates cutoffs with the potential. So, the 
liquid current peak could in principle be enlarged by increasing the energy separation between the rQW state (intermediate band) and the ground state of the lQW, responsible to uncouple
the ratchet state to the valence band. 

The saturation, which determines the liquid current peak intensity, is mainly drive by the first term at the right-hand side of~\ref{eqLiqdCur}. 
It takes into account the relationship between both the rQW recombination ($R_{iv}^r$) and generation ($G_{ic}^r$) rates, regarding to the 
transition rate between the wells ($T_i$) itself. Once more the recombination rate $R_{iv}^r$, with the same step-like shape and maximum intensity of around 100~GHz, 
is responsible for controls the liquid current intensity. Increasing $R_{iv}^r$ decreases the electrons concentration at the intermediate band and, consequently, the 
probability of such electrons to be scattered to the lQW, within rate $T_i$. 
Therefore, for $T_i$ lesser than such a recombination rate, we expected a lay back in the cell's current as observed.

Our results has been shown a clear enhancement in the generation of current by the intermediate band solar cell, within the excited state of the lQW 
acting as a quantum ratchet state. This is in consonance with Noda and coworkers~\cite{NODA16}, who developed a thermodynamical model and concluded the 
use of intermediate 
band solar cells, adding the ratchet state is expected to always enhance solar cell's efficiency beyond the Shockley-Quiesser 
limit~\cite{SCHO61}. They also showed that the inclusion of the intermediate band by itself is not a determining condition for increasing efficiency. The work provided
conditions where the intermediate band solar cells could be less efficient than singled junction ones. According to them,
an intermediate band solar cell will always be more efficient then a singled junction one, and overcome the Shockley-Queisser limit, only when adding ratchet states.

\begin{figure}
 \centering
 \includegraphics[scale=0.15]{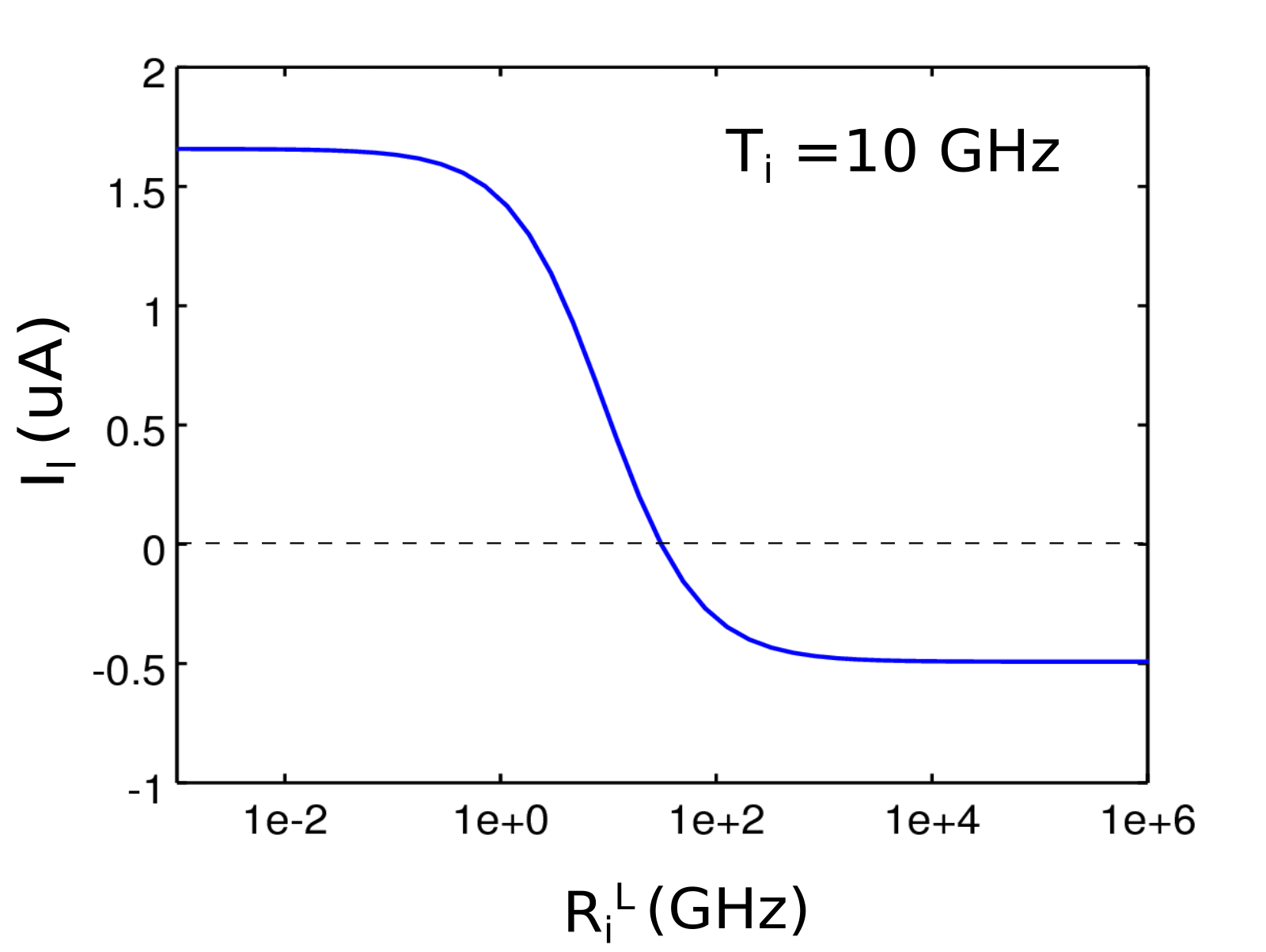}
 % RIL_T1d10.pdf: 0x0 pixel, 300dpi, 0.00x0.00 cm, bb=
 \caption{Liquid current as a function of the transition rate between the ground and excited subband at the rQW, $R_i^L$, for $T_i=10$~GHz and $3.8$~eV. We can observe a 
  current's signal change around $R_i^l = 70$~GHz.}
 \label{fig:RIL}
\end{figure}

However, we observed that even introducing the quantum ratchet state, such a behavior can be changed. As discussed earlier, the liquid current behavior is a consequence of the relationship of 
several generation and transition processes. ~\ref{fig:RIL}, presents the liquid current behavior, for $T_i = 10$~GHz and a backward potential of $3.8$~eV, as a 
function of the intersubband recombination rate $R_i^l$. We can observe the change in the liquid current signal, which becomes negative for higher $R_i^l$, 
indicating that increasing such a rate can be harmful to the liquid current for specific conditions. 

Independently of
how effective is the coupling between the intermediate band and the ratchet state, regarding of $T_i$ rate, our model shows there are conditions in which the introduction of the ratchet 
states is not a guarantee of increasing cell's current. Allowing electrons to be scattered into the ratchet state should increase their lifetime into the intermediate
band, but the relationship between the recombination and the scattering rates can both increase or decrease the solar cell current, depending on the other rates 
involved in the cell's dynamics. 

\section{Conclusion}
In summary, we have demonstrated the possibility of using the subbands of quantum wells as the intermediate band and ratchet state of an intermediate band 
solar cell with ratchet state, using a quantum cascade like approach. 
The cell was considered to be a double quantum well structure within GaAlAs and GaAs layers sandwiched between heavily doped GaAs
contacts. The system's eigenfunctions were numerically simulated by solving the time-dependent Schrödinger equation. The recombination and generation rates, used in 
a semiclassical rates model, were obtained by means of the Fermi golden rule, within the simulated eigenstates. 
With allowing electron to scattering inside the ratchet state, as a LO-phonon assisted scattering, we have shown an effective increase in the cell's current for specific values of electrical potential, 
close to the built-in potential of the cell.
Agreeing with other works, the inclusion of the ratchet 
state increases the solar cell current, by increasing the electron's lifetime into the intermediate band. However, the addition of the ratchet state is not the unique 
condition determining the cell's efficiency. The recombination and generation dynamics plays a fundamental role, and we can achieve situations in which the solar cell
current is decreased even with the inclusion of the ratchet state.

\begin{acknowledgments}
We thank CNPq and CAPES for research fellowships and for direct and indirect research funding. N.S. thanks CAPES for a Visiting Senior Professorship at UFABC.
The authors thank CNPq for providing funding for this research through INCT-DISSE consortium.
\end{acknowledgments}

\end{document}